\documentclass[aps,prc,superscriptaddress,showpacs,floatfix,twocolumn]{revtex4}

\usepackage{graphicx}
\usepackage[scriptsize]{caption}

\newcommand{\mean}[1]{\left\langle #1 \right\rangle} 

\newcommand{\pt}{$p_{\rm T}$}

\newcommand{\pp}{{p+p}\ }
\newcommand{\auau}{{\rm Au+Au\ }}

\newcommand{\qhatave}{$\langle$\^{q}$\rangle$}
\newcommand{\pvalue}{p-value}

\begin{document}

\title{Quantitative Constraints on the Transport Properties of Hot Partonic Matter from 
Semi-Inclusive Single High Transverse Momentum Pion Suppression in \auau
Collisions at $\sqrt{s_{NN}} = 200$~GeV.
}


\begin{abstract}

The PHENIX experiment has measured the suppression of semi-inclusive 
single high transverse momentum $\pi^{0}$'s in \auau collisions at 
$\sqrt{{s}_{NN}} = 200$ GeV.  The present understanding of this 
suppression is in terms of energy-loss of the parent (fragmenting) parton 
in a dense color-charge medium. 
%
%
We have performed a quantitative 
comparison between various parton energy-loss models and our experimental 
data.  The statistical point-to-point uncorrelated as well as correlated 
systematic uncertainties are taken into account in the comparison.  We 
detail this methodology and the resulting constraint on the model 
parameters, such as the initial 
color-charge density $dN^{g}/dy$, the medium transport coefficient \qhatave, or
the initial energy-loss parameter $\epsilon_{0}$. 
We find that high transverse momentum $\pi^{0}$ suppression in 
\auau collisions has sufficient precision to constrain these model 
dependent parameters at the $\pm$ 20--25\% (one standard deviation) 
level.  These constraints include only the experimental uncertainties, and 
further studies are needed to compute the corresponding theoretical 
uncertainties. 

\end{abstract}

\newcommand{\abilene}{Abilene Christian University, Abilene, TX 79699, USA}
\newcommand{\banaras}{Department of Physics, Banaras Hindu University, Varanasi 221005, India}
\newcommand{\bnl}{Brookhaven National Laboratory, Upton, NY 11973-5000, USA}
\newcommand{\caucr}{University of California - Riverside, Riverside, CA 92521, USA}
\newcommand{\charlesczech}{Charles University, Ovocn\'{y} trh 5, Praha 1, 116 36, Prague, Czech Republic}
\newcommand{\ciae}{China Institute of Atomic Energy (CIAE), Beijing, People's Republic of China}
\newcommand{\cns}{Center for Nuclear Study, Graduate School of Science, University of Tokyo, 7-3-1 Hongo, Bunkyo, Tokyo 113-0033, Japan}
\newcommand{\colorado}{University of Colorado, Boulder, CO 80309, USA}
\newcommand{\columbia}{Columbia University, New York, NY 10027 and Nevis Laboratories, Irvington, NY 10533, USA}
\newcommand{\czechtech}{Czech Technical University, Zikova 4, 166 36 Prague 6, Czech Republic}
\newcommand{\dapnia}{Dapnia, CEA Saclay, F-91191, Gif-sur-Yvette, France}
\newcommand{\debrecen}{Debrecen University, H-4010 Debrecen, Egyetem t{\'e}r 1, Hungary}
\newcommand{\elte}{ELTE, E{\"o}tv{\"o}s Lor{\'a}nd University, H - 1117 Budapest, P{\'a}zm{\'a}ny P. s. 1/A, Hungary}
\newcommand{\fit}{Florida Institute of Technology, Melbourne, FL 32901, USA}
\newcommand{\fsu}{Florida State University, Tallahassee, FL 32306, USA}
\newcommand{\gsu}{Georgia State University, Atlanta, GA 30303, USA}
\newcommand{\hiroshima}{Hiroshima University, Kagamiyama, Higashi-Hiroshima 739-8526, Japan}
\newcommand{\ihepprot}{IHEP Protvino, State Research Center of Russian Federation, Institute for High Energy Physics, Protvino, 142281, Russia}
\newcommand{\illuiuc}{University of Illinois at Urbana-Champaign, Urbana, IL 61801, USA}
\newcommand{\instpasczech}{Institute of Physics, Academy of Sciences of the Czech Republic, Na Slovance 2, 182 21 Prague 8, Czech Republic}
\newcommand{\isu}{Iowa State University, Ames, IA 50011, USA}
\newcommand{\jinrdubna}{Joint Institute for Nuclear Research, 141980 Dubna, Moscow Region, Russia}
\newcommand{\kaeri}{KAERI, Cyclotron Application Laboratory, Seoul, Korea}
\newcommand{\kek}{KEK, High Energy Accelerator Research Organization, Tsukuba, Ibaraki 305-0801, Japan}
\newcommand{\kfki}{KFKI Research Institute for Particle and Nuclear Physics of the Hungarian Academy of Sciences (MTA KFKI RMKI), H-1525 Budapest 114, POBox 49, Budapest, Hungary}
\newcommand{\korea}{Korea University, Seoul, 136-701, Korea}
\newcommand{\kurchatov}{Russian Research Center ``Kurchatov Institute", Moscow, Russia}
\newcommand{\kyoto}{Kyoto University, Kyoto 606-8502, Japan}
\newcommand{\labllr}{Laboratoire Leprince-Ringuet, Ecole Polytechnique, CNRS-IN2P3, Route de Saclay, F-91128, Palaiseau, France}
\newcommand{\lawllnl}{Lawrence Livermore National Laboratory, Livermore, CA 94550, USA}
\newcommand{\losalamos}{Los Alamos National Laboratory, Los Alamos, NM 87545, USA}
\newcommand{\lpc}{LPC, Universit{\'e} Blaise Pascal, CNRS-IN2P3, Clermont-Fd, 63177 Aubiere Cedex, France}
\newcommand{\lund}{Department of Physics, Lund University, Box 118, SE-221 00 Lund, Sweden}
\newcommand{\muenster}{Institut f\"ur Kernphysik, University of Muenster, D-48149 Muenster, Germany}
\newcommand{\myongji}{Myongji University, Yongin, Kyonggido 449-728, Korea}
\newcommand{\nagasaki}{Nagasaki Institute of Applied Science, Nagasaki-shi, Nagasaki 851-0193, Japan}
\newcommand{\newmex}{University of New Mexico, Albuquerque, NM 87131, USA }
\newcommand{\nmsu}{New Mexico State University, Las Cruces, NM 88003, USA}
\newcommand{\ornl}{Oak Ridge National Laboratory, Oak Ridge, TN 37831, USA}
\newcommand{\orsay}{IPN-Orsay, Universite Paris Sud, CNRS-IN2P3, BP1, F-91406, Orsay, France}
\newcommand{\peking}{Peking University, Beijing, People's Republic of China}
\newcommand{\pnpi}{PNPI, Petersburg Nuclear Physics Institute, Gatchina, Leningrad region, 188300, Russia}
\newcommand{\riken}{RIKEN, The Institute of Physical and Chemical Research, Wako, Saitama 351-0198, Japan}
\newcommand{\rikjrbrc}{RIKEN BNL Research Center, Brookhaven National Laboratory, Upton, NY 11973-5000, USA}
\newcommand{\rikkyo}{Physics Department, Rikkyo University, 3-34-1 Nishi-Ikebukuro, Toshima, Tokyo 171-8501, Japan}
\newcommand{\saispbstu}{Saint Petersburg State Polytechnic University, St. Petersburg, Russia}
\newcommand{\saopaulo}{Universidade de S{\~a}o Paulo, Instituto de F\'{\i}sica, Caixa Postal 66318, S{\~a}o Paulo CEP05315-970, Brazil}
\newcommand{\seoulnat}{System Electronics Laboratory, Seoul National University, Seoul, Korea}
\newcommand{\stonybrkc}{Chemistry Department, Stony Brook University, Stony Brook, SUNY, NY 11794-3400, USA}
\newcommand{\stonycrkp}{Department of Physics and Astronomy, Stony Brook University, SUNY, Stony Brook, NY 11794, USA}
\newcommand{\subatech}{SUBATECH (Ecole des Mines de Nantes, CNRS-IN2P3, Universit{\'e} de Nantes) BP 20722 - 44307, Nantes, France}
\newcommand{\tenn}{University of Tennessee, Knoxville, TN 37996, USA}
\newcommand{\titech}{Department of Physics, Tokyo Institute of Technology, Oh-okayama, Meguro, Tokyo 152-8551, Japan}
\newcommand{\tsukuba}{Institute of Physics, University of Tsukuba, Tsukuba, Ibaraki 305, Japan}
\newcommand{\vandy}{Vanderbilt University, Nashville, TN 37235, USA}
\newcommand{\waseda}{Waseda University, Advanced Research Institute for Science and Engineering, 17 Kikui-cho, Shinjuku-ku, Tokyo 162-0044, Japan}
\newcommand{\weizmann}{Weizmann Institute, Rehovot 76100, Israel}
\newcommand{\yonsei}{Yonsei University, IPAP, Seoul 120-749, Korea}
\affiliation{\abilene}
\affiliation{\banaras}
\affiliation{\bnl}
\affiliation{\caucr}
\affiliation{\charlesczech}
\affiliation{\ciae}
\affiliation{\cns}
\affiliation{\colorado}
\affiliation{\columbia}
\affiliation{\czechtech}
\affiliation{\dapnia}
\affiliation{\debrecen}
\affiliation{\elte}
\affiliation{\fit}
\affiliation{\fsu}
\affiliation{\gsu}
\affiliation{\hiroshima}
\affiliation{\ihepprot}
\affiliation{\illuiuc}
\affiliation{\instpasczech}
\affiliation{\isu}
\affiliation{\jinrdubna}
\affiliation{\kaeri}
\affiliation{\kek}
\affiliation{\kfki}
\affiliation{\korea}
\affiliation{\kurchatov}
\affiliation{\kyoto}
\affiliation{\labllr}
\affiliation{\lawllnl}
\affiliation{\losalamos}
\affiliation{\lpc}
\affiliation{\lund}
\affiliation{\muenster}
\affiliation{\myongji}
\affiliation{\nagasaki}
\affiliation{\newmex}
\affiliation{\nmsu}
\affiliation{\ornl}
\affiliation{\orsay}
\affiliation{\peking}
\affiliation{\pnpi}
\affiliation{\riken}
\affiliation{\rikjrbrc}
\affiliation{\rikkyo}
\affiliation{\saispbstu}
\affiliation{\saopaulo}
\affiliation{\seoulnat}
\affiliation{\stonybrkc}
\affiliation{\stonycrkp}
\affiliation{\subatech}
\affiliation{\tenn}
\affiliation{\titech}
\affiliation{\tsukuba}
\affiliation{\vandy}
\affiliation{\waseda}
\affiliation{\weizmann}
\affiliation{\yonsei}
\author{A.~Adare}	\affiliation{\colorado}
\author{S.~Afanasiev}	\affiliation{\jinrdubna}
\author{C.~Aidala}	\affiliation{\columbia}
\author{N.N.~Ajitanand}	\affiliation{\stonybrkc}
\author{Y.~Akiba}	\affiliation{\riken} \affiliation{\rikjrbrc}
\author{H.~Al-Bataineh}	\affiliation{\nmsu}
\author{J.~Alexander}	\affiliation{\stonybrkc}
\author{A.~Al-Jamel}	\affiliation{\nmsu}
\author{K.~Aoki}	\affiliation{\kyoto} \affiliation{\riken}
\author{L.~Aphecetche}	\affiliation{\subatech}
% ...
\collaboration{PHENIX Collaboration} \noaffiliation


\maketitle


\section{Introduction}

Heavy ion collisions at very high energy are of interest due to the formation of a novel
partonic medium approximately the size of a large nucleus, but with an energy density exceeding
that of normal nuclei by considerably more than an order of magnitude.  At such high energy densities, it is 
believed that quarks and gluons are no longer confined in hadrons, but may be constituents of a 
quark-gluon plasma with characteristics of a near-perfect fluid~(for a detailed review see \cite{nagle_review_articles}).  Experiments at the 
Relativistic Heavy Ion Collider (RHIC) have already demonstrated that a very hot and dense, strongly interacting medium
is created in \auau collisions at $\sqrt{{s}_{NN}}=200$ GeV~\cite{RHIC_whitepapers}.
The goal is now to quantitatively determine the properties of this medium.

Important properties of the medium include the density of color-charges as well
as the exchange of transverse momentum between parton probes and the medium.
In rare events, in addition to the creation of the medium, there can also be a hard
scattering (high-$Q^2$ process) between the colliding partons that, 
at leading order, sends two high-energy quark or
gluon partons in opposite transverse directions.  These high-energy partons can be utilized to probe
both the color-charge density of the medium and the coupling strength between the parton
and the medium.  
There are various calculational frameworks for modeling these interactions (for a detailed review see \cite{elossreview}).

In this paper, we consider four specific calculations of parton energy-loss (discussed below):
the Parton Quenching Model (PQM)~\cite{PQM}, the Gyulassy-Levai-Vitev (GLV) model~\cite{GLV},
the Wicks-Horowitz-Djordjevic-Gyulassy (WHDG) model~\cite{Horowitz}, and the 
Zhang-Owens-Wang-Wang (ZOWW) model~\cite{zoww}.
We detail a quantitative method of assessing the sensitivity of the latest
measurements to the input parameters of these models that characterize the initial parton density
or medium transport coefficients.


\section{Experimental Results}

During the 2004 data-taking period at the Relativistic Heavy Ion 
Collider, the PHENIX experiment recorded an integrated luminosity of 0.24 
nb$^{-1}$ in $\sqrt{s_\mathrm{NN}}$=200~GeV \auau collisions, which 
extends the measurement of $\pi^0$ to much higher \pt\ than previous data 
sets allowed. The results and further details of this measurement are 
given in~\cite{ppg080}. A brief description is given below.

The PHENIX experiment measures $\pi^0$'s via the two-photon decay mode
with two types of highly segmented ($\Delta\eta \times \Delta\phi
\approx 0.01 \times 0.01$) electromagnetic calorimeters~(EMCal) located
at the radial distance of approximately 5~m from the vertex~\cite{EMCal}.
One is a lead scintillator sampling calorimeter~(PbSc), which covers the
geometrical acceptance of $|\eta|<0.35$ and $\Delta\phi = 3/4 \pi$.
The other is a lead glass Cerenkov calorimeter~(PbGl), whose geometrical
coverage is $|\eta|<0.35$ and $\Delta\phi = \pi/4$. 
The energy resolution of the PbSc and PbGl calorimeters as determined
from test beam measurements are given by
$8.1\%/\sqrt{E(\mathrm{GeV})} \oplus 2.1\%$ and
$5.9\%/\sqrt{E(\mathrm{GeV})} \oplus 0.8\%$, respectively. 
The energy calibration of the EMCal modules is based upon the
measured position of the $\pi^0$ mass peak, the deposited energy of
minimum ionizing particles which traverse the calorimeter, and the ratio of energy
to momentum which is expected to be about 1 for electrons identified
by the Ring-Imaging Cerenkov detector.
The systematic uncertainty on the energy scale is $\sim$ 1~\%, which
corresponds to $\sim$ 7-12~\% uncertainty on the invariant $\pi^0$ yield
over the \pt\ range of the measurement.

Neutral pions were reconstructed in their $\pi^{0} \rightarrow \gamma \gamma$  
decay channel.  
Photon candidates are identified by applying particle identification cuts
based mainly on the shower shape.  The invariant mass for all photon-pair
combinations within one event that satisfy cuts on the energy
asymmetry $|E_{\gamma1}-E_{\gamma2}|/|E_{\gamma1}+E_{\gamma2}|<0.8$ were
calculated in bins of \pt.  
The combinatorial background is determined by combining into pairs uncorrelated
photons from different events with similar centrality, reaction
plane, and vertex location. 

\begin{figure}[tbh]
\includegraphics[width=1.0\linewidth]{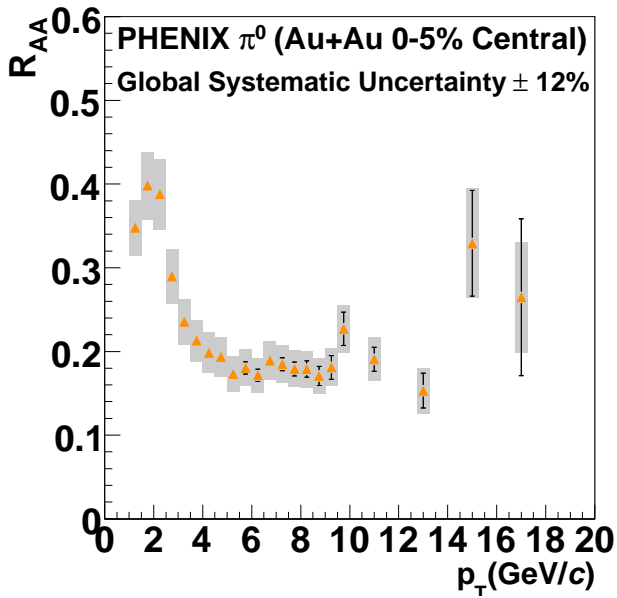}
\caption[]{
(Color online) The $\pi^{0}$ nuclear suppression factor $R_{\rm AA}$ as a function of transverse
momentum for 0-5\% \auau collisions at $\sqrt{s_{NN}}$=200 GeV.  Point-to-point uncorrelated statistical
and systematic uncertainties are shown as uncertainty bars.  
Correlated systematic uncertainties are shown as gray boxes
around the data points.  The global scale factor systematic uncertainty is $\pm$12\%.
}
\label{fig_data_only}
\end{figure}

The raw $\pi^0$ yield was obtained by integrating the mass peak region
of the invariant mass distribution after subtracting the combinatorial
background.
The raw spectra are corrected for the detector response (energy
resolution), the reconstruction efficiency, and occupancy effects
(e.g. overlapping clusters).
These corrections are made by embedding simulated single $\pi^0$'s from a
full GEANT simulation of the PHENIX detector into real events, and analyzing the embedded
$\pi^0$ events with the same analysis cuts as used with real events.

After computing the invariant yields in \auau collisions~\cite{ppg080}, 
the medium effects are quantified using the nuclear modification
factor~($R_{\rm AA}$). $R_{\rm AA}$ is the ratio between the measured yield
and the expected yield for point-like processes 
scaled from the \pp result, and is defined as:

\begin{eqnarray}
 \mathrm{R_{\rm AA}}(p_\mathrm{T}) =
  \frac{(1/N^{evt}_{\rm AA}) d^2N^{\pi^{0}}_\mathrm{\rm AA}/dp_\mathrm{T}d y}{{\langle T_\mathrm{\rm AA} \rangle} d^2\sigma^{\pi^{0}}_\mathrm{NN}/dp_\mathrm{T}d y} \qquad ,
\end{eqnarray}
where~${\langle T_\mathrm{\rm AA}\rangle}$ is the average Glauber nuclear overlap
function for the \auau centrality bin under consideration
\begin{equation}
{\langle T_{\rm AA}\rangle}\equiv
\frac {\int T_{\rm AA}({\bf b}) \, d{\bf b} }{\int (1- e^{- 
\sigma^{\rm inel}_{NN}\, T_{\rm AA}({\bf b})})\, d{\bf b}}=
\langle N_{\rm coll}^{\sigma_{NN}}\rangle/\sigma^{\rm inel}_{NN} \quad.
\label{eq:RAA}
\end{equation}
where $\langle N_{\rm coll}^{\sigma_{NN}}\rangle$ is the average  
number of inelastic nucleon-nucleon collisions for the \auau centrality bin under
consideration calculated with inelastic nucleon-nucleon cross section 
$\sigma^{\rm inel}_{NN}$.

\begin{table*}[htbp]
\caption{\label{tab:data}
The $\pi^{0}$ nuclear suppression factor $R_{\rm AA}$ as a function of transverse
momentum for 0-5\% \auau collisions at $\sqrt{s_{NN}}$=200 GeV.  The type A, B, 
and C uncertainties are tabulated for each point.}
\begin{ruledtabular} \begin{tabular}{ccccc}
~$p_{\rm T}$ (GeV/$c$)~ & ~~~$R_{\rm AA}$~~~ & ~Type A Uncertainty~ & ~Type B Uncertainty~ & ~Type C Uncertainty~ \\
\hline
1.25 & 0.347 & $\pm 0.007 $& $\pm 0.033 $& $\pm  0.041 $ \\
1.75 & 0.398 & $\pm 0.007 $& $\pm 0.040 $& $\pm  0.047 $ \\
2.25 & 0.387 & $\pm 0.007 $& $\pm 0.042 $& $\pm  0.046 $ \\
2.75 & 0.289 & $\pm 0.006 $& $\pm 0.032 $& $\pm  0.034 $ \\
3.25 & 0.235 & $\pm 0.005 $& $\pm 0.027 $& $\pm  0.028 $ \\
3.75 & 0.21  & $\pm 0.005 $& $\pm 0.024 $& $\pm  0.025 $ \\
4.25 & 0.198 & $\pm 0.005 $& $\pm 0.024 $& $\pm  0.023 $ \\
4.75 & 0.193 & $\pm 0.006 $& $\pm 0.023 $& $\pm  0.023 $ \\
5.25 & 0.172 & $\pm 0.006 $& $\pm 0.021 $& $\pm  0.020 $ \\
5.75 & 0.180 & $\pm 0.007 $& $\pm 0.021 $& $\pm  0.021 $ \\
6.25 & 0.171 & $\pm 0.007 $& $\pm 0.020 $& $\pm  0.020 $ \\
6.75 & 0.189 & $\pm 0.007 $& $\pm 0.022 $& $\pm  0.022 $ \\
7.25 & 0.184 & $\pm 0.008 $& $\pm 0.022 $& $\pm  0.022 $ \\
7.75 & 0.179 & $\pm 0.008 $& $\pm 0.021 $& $\pm  0.021 $ \\
8.25 & 0.178 & $\pm 0.010 $& $\pm 0.021 $& $\pm  0.021 $ \\
8.75 & 0.170 & $\pm 0.011 $& $\pm 0.020 $& $\pm  0.020 $ \\
9.25 & 0.180 & $\pm 0.014 $& $\pm 0.022 $& $\pm  0.021 $ \\
9.75 & 0.226 & $\pm 0.019 $& $\pm 0.028 $& $\pm  0.027 $ \\
11.00 & 0.190 & $\pm 0.014 $& $\pm 0.026 $& $\pm 0.022 $ \\
13.00 & 0.153 & $\pm 0.020 $& $\pm 0.027 $& $\pm  0.018 $ \\
15.00 & 0.329 & $\pm 0.063 $& $\pm 0.065 $& $\pm  0.039 $ \\
17.00 & 0.264 & $\pm 0.093 $& $\pm 0.065 $& $\pm  0.031 $ \\
\end{tabular} \end{ruledtabular}
\end{table*}

The measured nuclear modification factors for 0-5\% central \auau 
reactions at $\sqrt{s_{NN}}=200$ GeV are shown in Fig.~\ref{fig_data_only} 
and tabulated in Table~\ref{tab:data} with statistical and systematic 
uncertainties. The systematic uncertainties fall into three categories. We 
denote Type A uncertainties as systematic uncertainties which are 
uncorrelated from point-to-point.  Type A systematic uncertainties are 
added in quadrature with statistical uncertainties and are shown as uncertainty
bars.  Partially correlated point-to-point systematic uncertainties are 
broken into a 100\% correlated component, referred to as Type B, and the 
above mentioned Type A.  The Type B uncertainties are shown as gray boxes.  
The sources of type B uncertainties are discussed in detail in 
~\cite{ppg080}, and are dominated by energy scale uncertainties, but also have 
contributions from photon shower merging at the highest \pt\ ($\approx 
15-20$ GeV/$c$).  There are also global systematic uncertainties, referred 
to as type C uncertainties, that are globally correlated systematic 
uncertainties (i.e. where all data points move by the same multiplicative 
factor).  The type C uncertainties are $\pm 12$\% and derive from 
uncertainties in the calculated nuclear thickness function and from the 
\pp\ absolute normalization.

\section{Theoretical Calculations}


The Parton Quenching Model (PQM)~\cite{PQM} encodes the dissipative
properties of the system in terms of a single transport-coefficient, often
referred to as \^{q}, obtained as the product of the parton-medium
cross-section times the color-charge density.
The average \^{q}~ quantifies the average squared transverse 
momentum transferred from the medium to the parton per mean free path.
The PQM model is a Monte Carlo program constructed using the 
quenching weights from BDMPS~\cite{BDMPS,salgado}.  
BDMPS is a perturbative calculation explicitly including only coherent radiative energy-loss
for the parton via gluon bremsstrahlung.  The PQM model incorporates a realistic transverse collision geometry,
though with a static medium.  
It is also notable that the PQM model
does not include initial state multiple scattering or modified nuclear parton distribution functions (PDF's). 

The Gyulassy-Levai-Vitev (GLV)~\cite{GLV} model is a formalism developed to calculate in-medium gluon
bremsstrahlung.  An analytic expression is derived for the single gluon emission spectrum to all orders in 
opacity, assuming an infrared cut-off given by the plasmon frequency.  Thus, within this framework, one can
extract the local color-charge density.
The color-charge density is written simply as  $dN^{g}/dy$, assuming a completely gluonic medium, or an equivalent $dN^{q,g}/dy$ for
a mixture of quarks and gluons.  In~\cite{Vitev}, using a realistic transverse collision geometry, the authors calculate {\it a priori},
without energy-loss, the single fixed geometrical average path length from the production point to the medium edge, 
$\langle L \rangle _{prod}$, and use it to calculate the parton energy-loss in a Bjorken expanding medium~\cite{bjorken}.
The calculation also incorporates initial state multiple scattering effects and modified nuclear PDF's.

The Wicks-Horowitz-Djordjevic-Gyulassy (WHDG) \cite{Horowitz} model utilizes the 
generalized GLV formalism~\cite{Djordjevic} for radiative
energy-loss described above.  In addition, their calculation includes a convolution of radiative energy-loss and
collisional energy-loss mechanisms.  A realistic transverse collision geometry with a 
Bjorken time expansion is utilized, and then a full distribution of parton paths through the medium is calculated.  The WHDG model does not
yet include initial state multiple scattering or modified PDF's.

The Zhang-Owens-Wang-Wang (ZOWW) \cite{zoww} calculations incorporate a next-to-leading order perturbative QCD parton
model with modified jet fragmentation functions.  The calculation explicitly includes only radiative energy-loss.
A hard-sphere transverse collision geometry with a one dimensional expanding medium is utilized.  The calculation
also incorporates initial state multiple scattering effects and modified nuclear PDF's.  

\begin{figure*}[tb]
\includegraphics[width=0.7\linewidth]{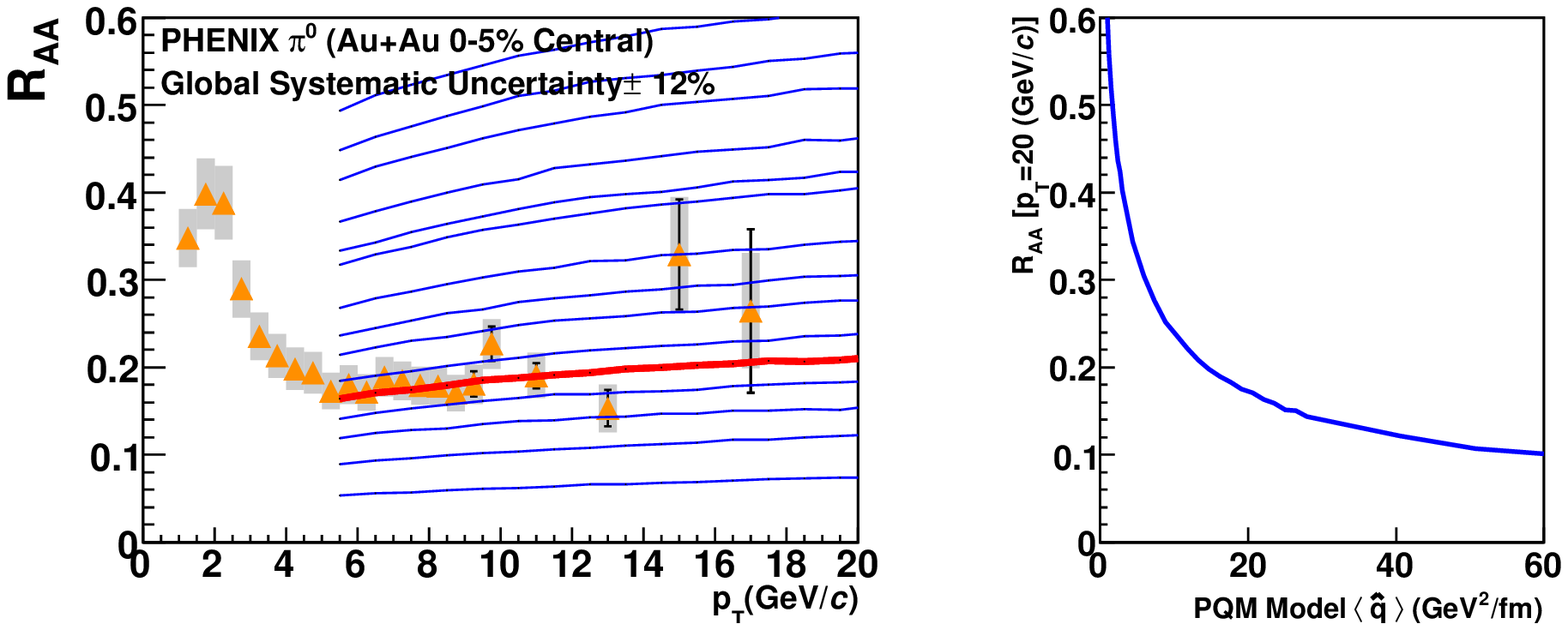}
\includegraphics[width=0.7\linewidth]{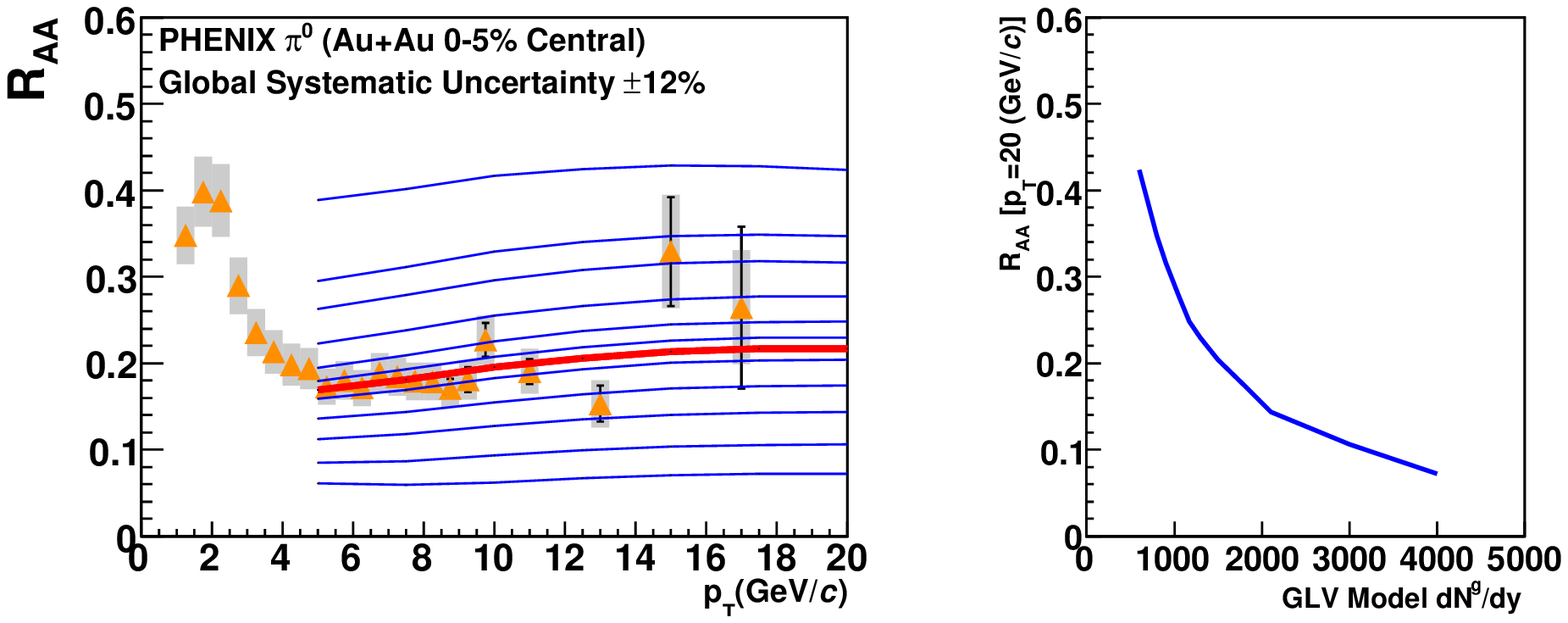}
\includegraphics[width=0.7\linewidth]{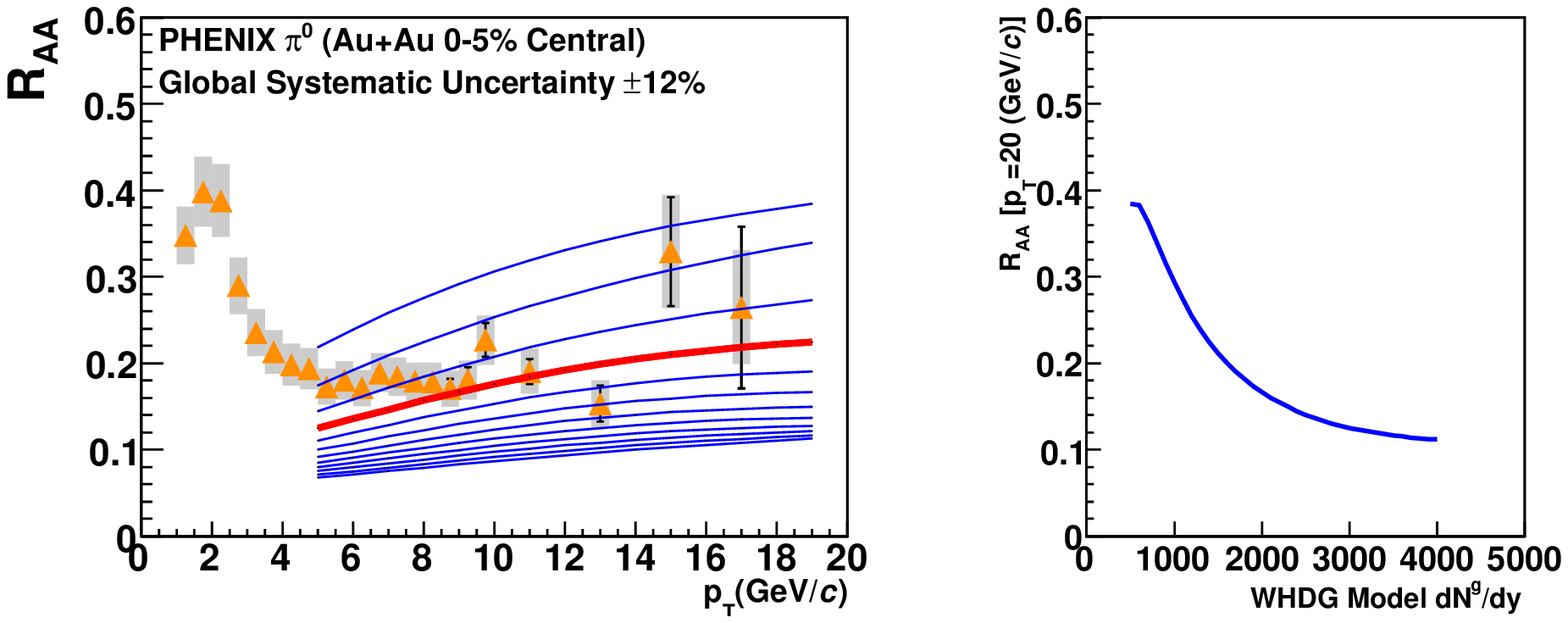}
\includegraphics[width=0.75\linewidth]{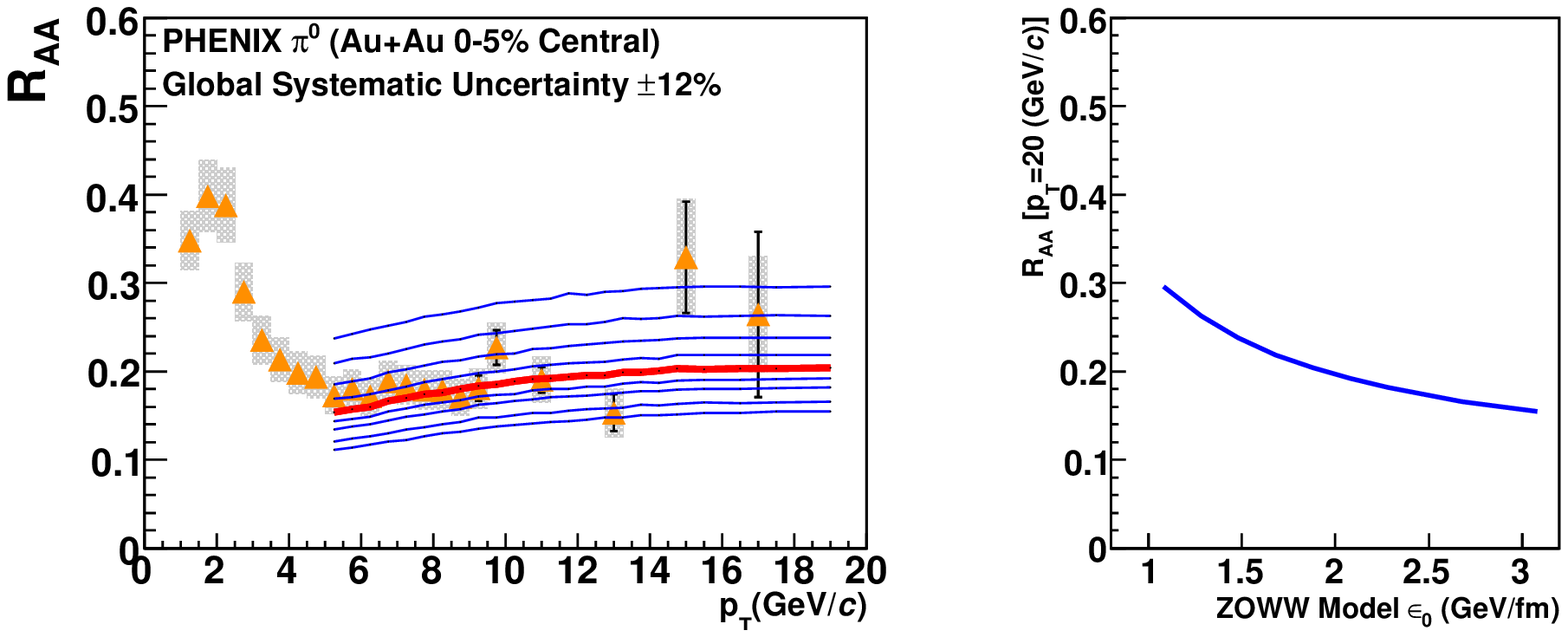}
\caption[]{\label{fig_data_models}
(Color online) Left panels show $\pi^{0}$ $R_{\rm AA}$ 
for 0-5\% \auau collisions at $\sqrt{s_{NN}}$=200 GeV and 
predictions from PQM~\protect\cite{PQM},
GLV~\protect\cite{Vitev},
WHDG~\protect\cite{Horowitz}, and ZOWW~\protect\cite{zoww} models
with (from top to bottom)
\qhatave~values of 0.3, 0.9, 1.2, 1.5, 2.1, 2.9, 4.4, 5.9, 7.4, 10.3, 13.2, 17.7, 25.0, 40.5, 101.4 GeV$^{2}$/fm;
$dN^{g}/dy$ values of 600, 800, 900, 1050, 1175, 1300, 1400, 1500, 1800, 2100, 3000, 4000; 
$dN^{g}/dy$ values of 500, 800, 1100, 1400, 1700, 2000, 2300, 2600, 2900, 3200, 3500, 3800; and
$\epsilon_{0}$ values of 1.08, 1.28, 1.48, 1.68, 1.88, 2.08, 2.28, 2.68, 3.08 GeV/fm.
Red lines indicate the best fit cases of 
(top) \qhatave = 13.2, 
(upper middle) $dN^{g}/dy$ = 1400,
(lower middle) $dN^{g}/dy$ = 1400,
and (bottom) $\epsilon_{0}$ = 1.88 GeV/fm.
Right panels show $R_{\rm AA}$ at $p_{\rm T}=20$ GeV/$c$.
}
\end{figure*}

The top left panel of Fig.~\ref{fig_data_models} shows a comparison of the experimental 
data with calculated results from the PQM energy-loss model (as described in~\cite{PQM})
corresponding to different \qhatave~ values~\cite{PQM_private}.
Note that only a subset of all the calculations corresponding to different \qhatave~values are
shown in the figure for clarity.
The upper right panel shows the $\pi^{0}$ suppression factor predicted at $p_{\rm T}=20$ GeV/$c$ 
from the PQM model as a function of the \qhatave~value.  
One can see that as the \qhatave~increases, the additional suppression becomes smaller (i.e. saturates).
This saturation effect was noted in~\cite{urs}, and interpretted as a result of the dominance of
preferential surface emission.

The other panels of Fig.~\ref{fig_data_models} show similar comparisons of the experimental 
data with calculated results utilizing the GLV numerical calculation framework (as described in~\cite{Vitev}) 
corresponding to different $dN^{g}/dy$ values~\cite{Vitev_private}; 
the WHDG calculational framework (as described in~\cite{Horowitz}) corresponding to different
$dN^{g}/dy$ values~\cite{Horowitz_private}; and 
the ZOWW calculational framework (as detailed in~\cite{zoww}) corresponding to different
$\epsilon_{0}$ values~\cite{zoww_private}.
Note that all calculations are shown only for \pt $> 5$ GeV/$c$ as that is 
where the calculations are considered applicable.

\begin{figure}[tbh]
\includegraphics[width=1.0\linewidth]{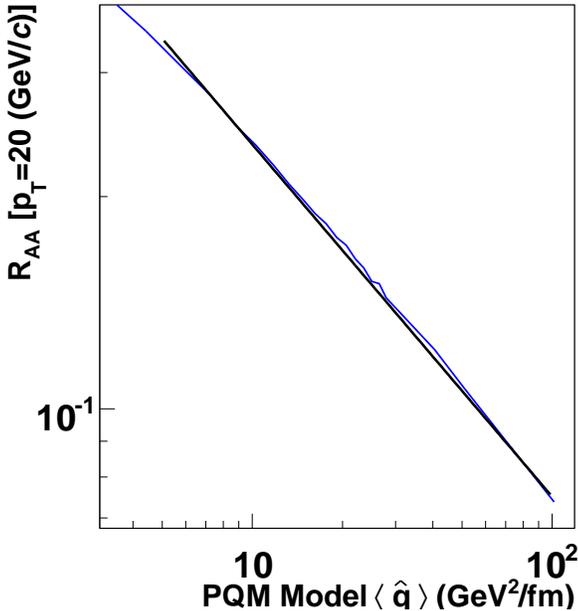}
\caption[]{
(Color online)  The nuclear suppression factors at $p_{\rm{T}}=20$ GeV/$c$
for PQM as a function of \qhatave are shown as a blue line with a log-x and
log-y display.  Also shown is the functional form $\Delta$ \qhatave
/\qhatave~$\approx$~2.0 $\cdot~\Delta R_{\rm{AA}}/R_{\rm{AA}}$) over the
range $5 <$ \qhatave~$< 100$ GeV$^{2}$/fm.
}
\label{fig_pqm_loglog}
\end{figure}

In Fig.~\ref{fig_pqm_loglog}, the same $\pi^{0}$ suppression factor
predicted at \pt = 20~GeV/$c$ from the PQM model as a function of the
\qhatave~value is shown, but in this case with a log-x and log-y scale. The
$R_{\rm{AA}} \approx 0.75/\sqrt{\langle \hat{q} \rangle}$ with \qhatave~in
units of GeV$^{2}$/fm over the range $5 <$ \qhatave~$< 100$. This means that
over this range for a given fractional change in \qhatave~there is always
the same fractional change in $R_{\rm{AA}}$ (i.e. $\Delta$ \qhatave
/\qhatave~$\approx$~2.0 $\cdot~\Delta R_{\rm{AA}}/R_{\rm{AA}}$).

\section{Combined Results}

\begin{figure}[tb]
\includegraphics[width=0.7\linewidth]{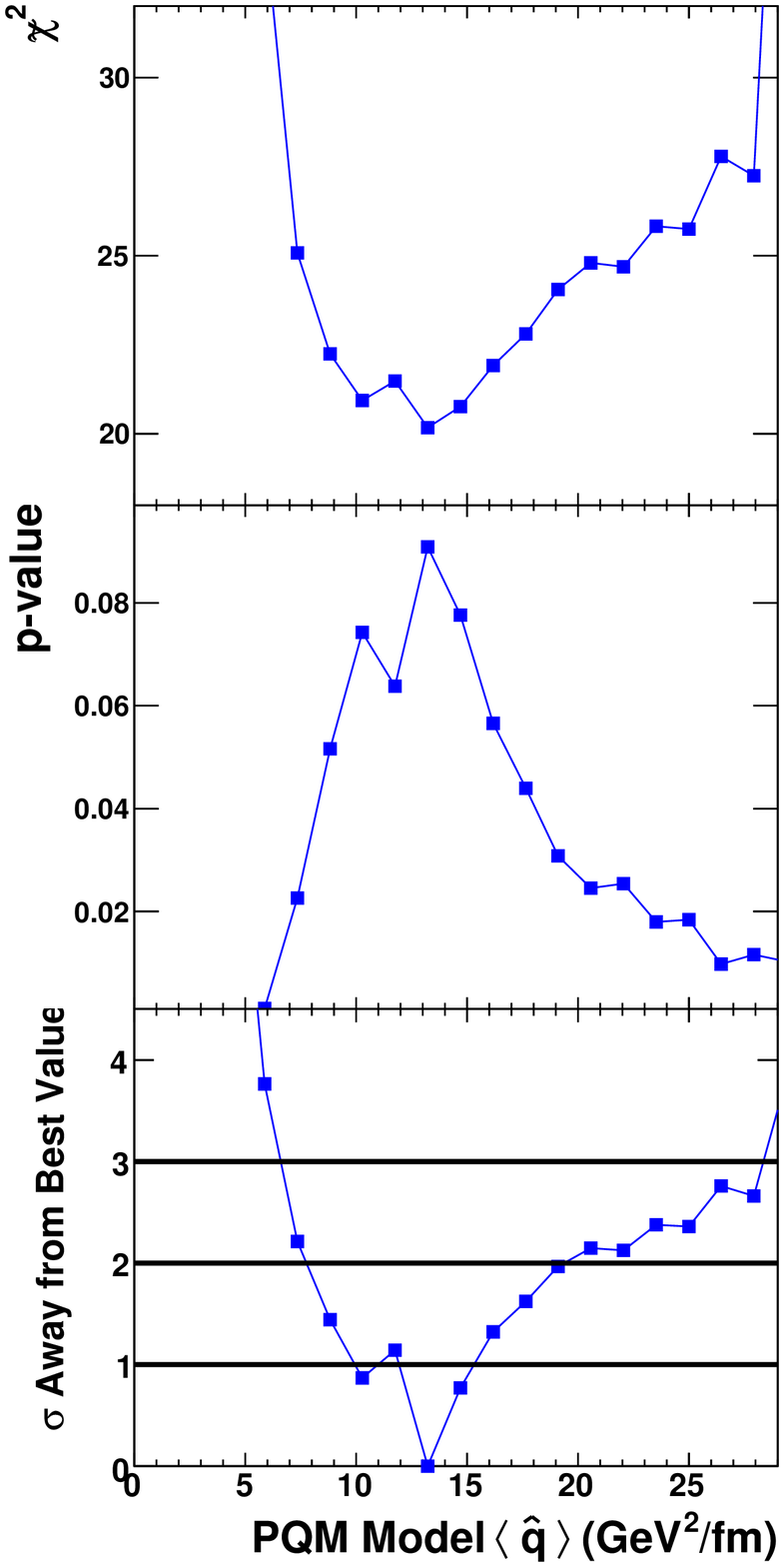}
\caption[]{
(Color online) The statistical analysis results from the comparison of the PQM model with 
the $\pi^{0}$ $R_{\rm AA}$(\pt) experimental data.  The top panel shows the 
modified $\tilde{\chi}^2$ for different values of the PQM \qhatave.  The 
middle panel shows the computed \pvalue~directly from the modified 
$\tilde{\chi}^2$ as shown above. The bottom panel shows the number of 
standard deviations ($\sigma$) away from the minimum (best) \qhatave~ parameter value 
for the PQM model calculations.
}
\label{fig_pqm_results}
\end{figure}

\begin{figure}[tb]
\includegraphics[width=0.7\linewidth]{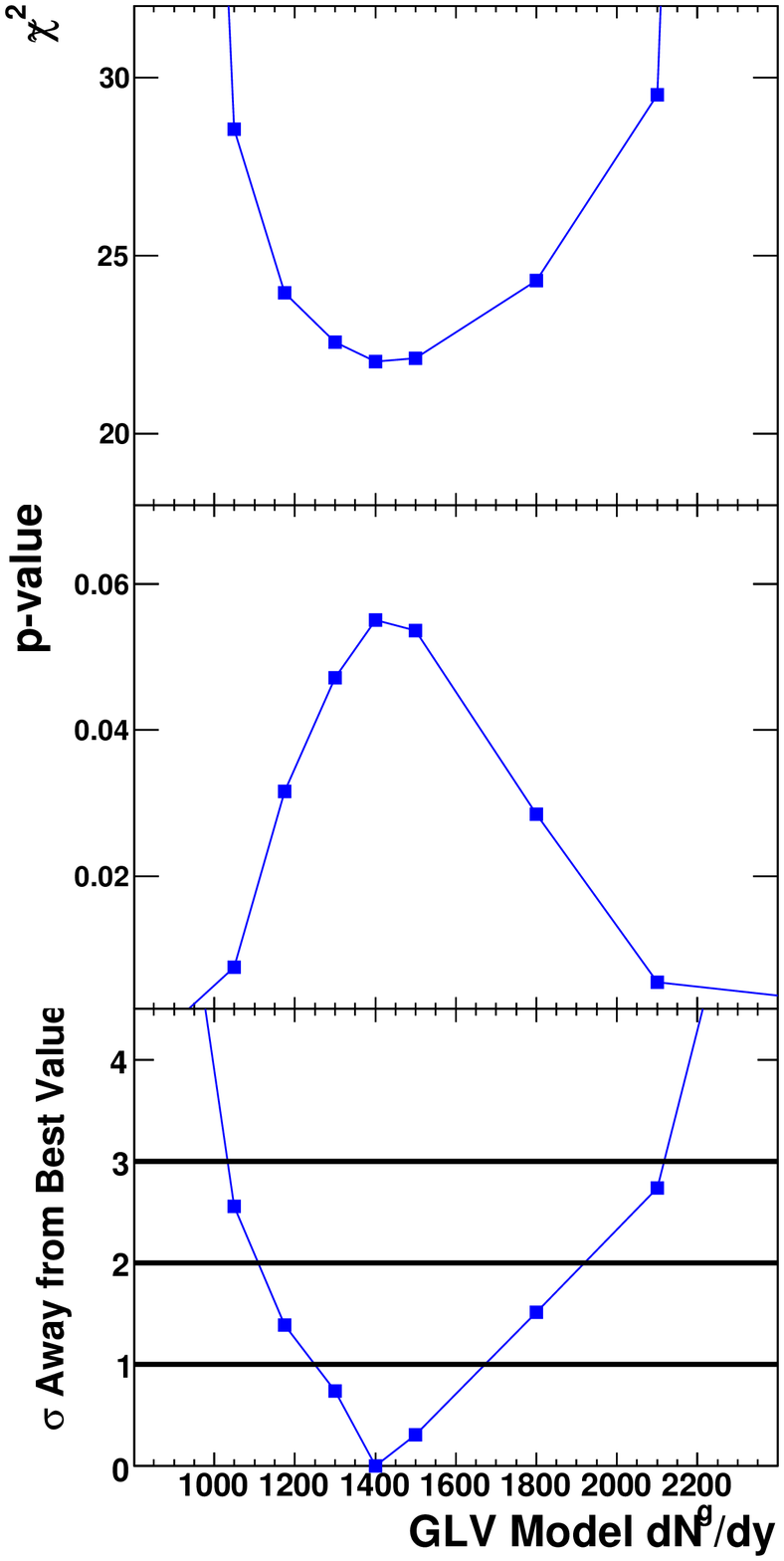}
\caption[]{
(Color online) The statistical analysis results from the comparison of the GLV model with the
$\pi^{0}$ $R_{\rm AA}$(\pt) experimental data.  The top panel shows the modified $\tilde{\chi}^2$
for different values of the GLV  $dN^{g}/dy$.  The middle panel shows the computed 
\pvalue~directly from the modified $\tilde{\chi}^2$ as shown above.
The bottom panel shows the number of standard deviations ($\sigma$) away from the minimum (best)  $dN^{g}/dy$
parameter value for the GLV model calculations.
}
\label{fig_glv_results}
\end{figure}

\begin{figure}[tb]
\includegraphics[width=0.7\linewidth]{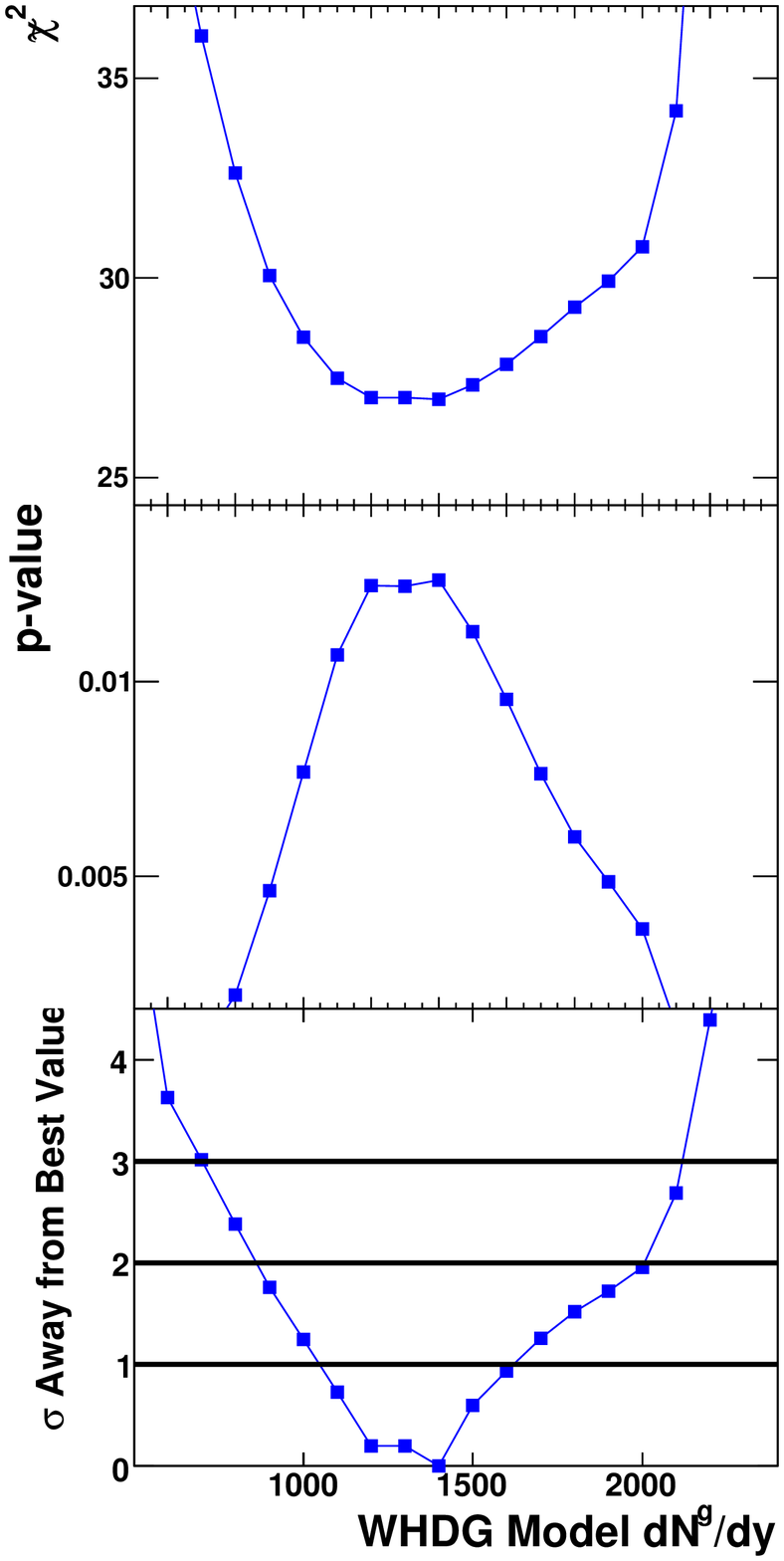}
\caption[]{
(Color online) The statistical analysis results from the comparison of the WHDG model with the
$\pi^{0}$ $R_{\rm AA}$(\pt) experimental data.  The top panel shows the modified $\tilde{\chi}^2$
for different values of the WHDG  $dN^{g}/dy$.  The middle panel shows the computed 
\pvalue~directly from the modified $\tilde{\chi}^2$ as shown above.
The bottom panel shows the number of standard deviations ($\sigma$) away from the minimum (best)  $dN^{g}/dy$
parameter value for the WHDG model calculations.
}
\label{fig_whdg_results}
\end{figure}

\begin{figure}[tb]
\includegraphics[width=0.7\linewidth]{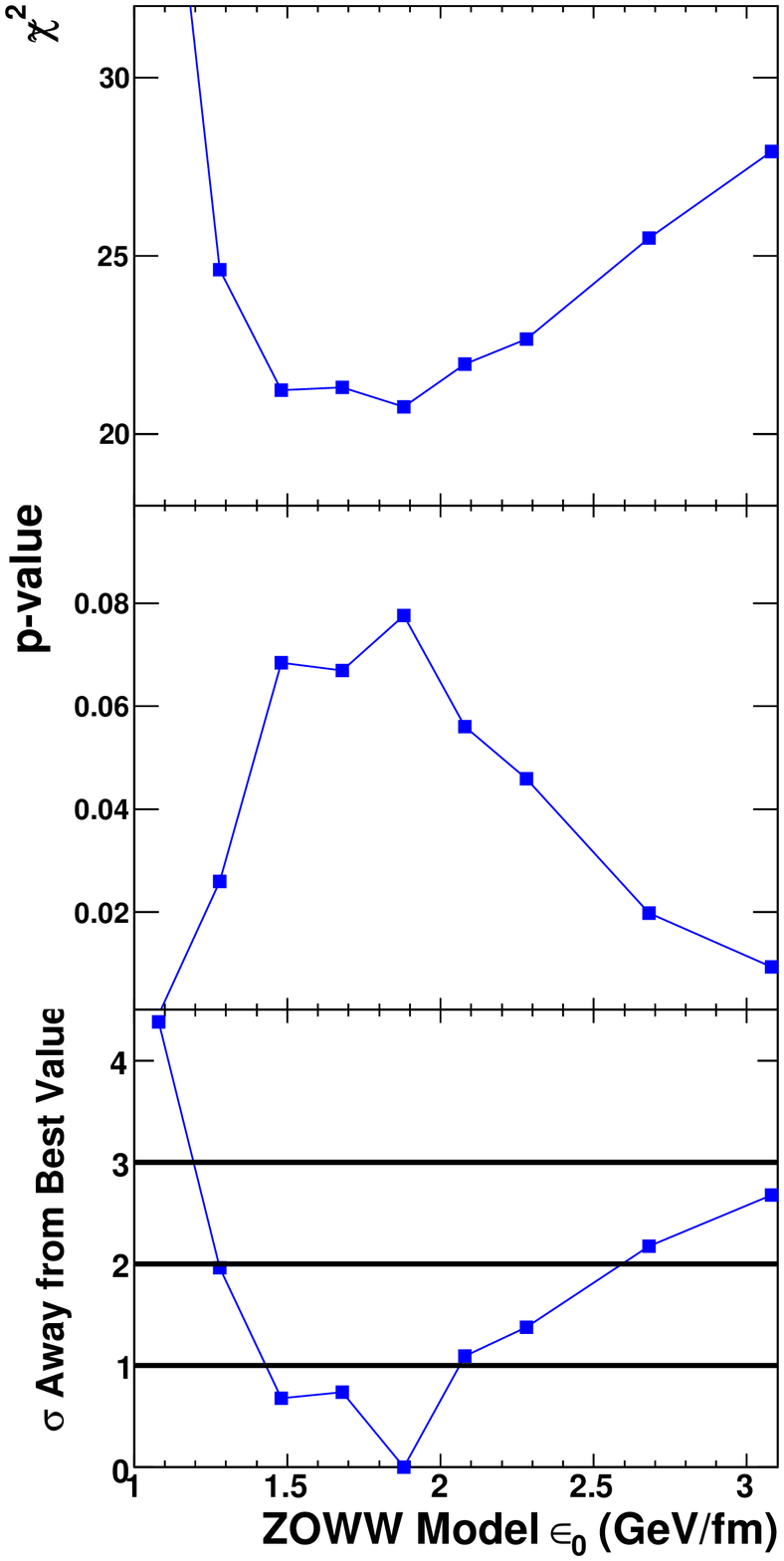}
\caption[]{
(Color online) The statistical analysis results from the comparison of the ZOWW model with the
$\pi^{0}$ $R_{\rm AA}$(\pt) experimental data.  The top panel shows the modified $\tilde{\chi}^2$
for different values of the ZOWW  $\epsilon_{0}$.  The middle panel shows the computed 
\pvalue~directly from the modified $\tilde{\chi}^2$ as shown above.
The bottom panel shows the number of standard deviations ($\sigma$) away from the minimum (best)  $\epsilon_{0}$
parameter value for the ZOWW model calculations.
}
\label{fig_zoww_results}
\end{figure}


The task is now to detail how the experimental uncertainties (statistical and systematic) 
constrain the model parameters 
that are reflected in the suppression factors.  
As described previously, the uncertainties of the measured points are separated into Type A ($p_{\rm T}$-uncorrelated, statistical $\oplus$ systematic, $\sigma_i$), Type B ($p_{\rm T}$-correlated, $\sigma_{b_i}$, boxes on Fig.~1), Type C (normalization, uniform fractional shift for all points, $\sigma_c$), where the $\sigma$'s represent the standard deviations of the assumed Gaussian distributed uncertainties. 
With the predicted theory value $\mu_i(p)$ for each data point calculated for different values of the input parameter $p$, 
we perform a least-squares fit the to the theory by finding the values of ${p}$, ${\epsilon}_b$, ${\epsilon}_c$ that minimize:   
\begin{widetext}
\begin{equation}
\tilde{\chi}^2(\epsilon_{b}, \epsilon_{c}, {p})={\left[ \left( \sum_{i=1}^{n}
{{(y_i+\epsilon_b \sigma_{b_i} +\epsilon_c y_i \sigma_c -\mu_i({p}))^2}  \over {\tilde{\sigma}_i^2}} \right) + {\epsilon_b^2 }+{\epsilon_c^2 }\right]} \qquad ,
\label{eq:-2lnL-gaussian-sys}
\end{equation}
\end{widetext}
where $\epsilon_b$ and $\epsilon_c$ are the fractions of the type B and C systematic uncertainties that all points are displaced together and where  
$\tilde{\sigma}_i=\sigma_i ({y_i+\epsilon_b \sigma_{b_i} +\epsilon_c y_i \sigma_c})/{y_i}$ is the point-to-point random uncertainty scaled by the multiplicative shift in $y_i$ such that the fractional uncertainty is unchanged under systematic shifts, which is true for the present measurement. 
For clarity of presentation, the derivation of Eq.~\ref{eq:-2lnL-gaussian-sys} (above) is given in Appendix A.

For any fixed values of $\epsilon_b$, $\epsilon_c$, Eq.~\ref{eq:-2lnL-gaussian-sys} follows the $\chi^2$ distribution with $n+2$ degrees of freedom, for testing the theoretical predictions $\mu_i(p)$, because it is the sum of $n+2$ Gaussian distributed random variables. The best fit, $\tilde{\chi}^2_{\rm min}$, the minimum of $\tilde{\chi}^2(\epsilon_{b}, \epsilon_{c}, {p})$ by variation of $\epsilon_{b}, \epsilon_{c},$ and ${p}$, is found by standard methods (for example using a MINUIT type minimization algorithm) and should follow the $\chi^2$ distribution with $n-1$ degrees of freedom.  
The correlated uncertainties of the best fit parameters are estimated in the Gaussian approximation by $\tilde{\chi}^2(\epsilon_{b}, \epsilon_{c}, {p})=\tilde{\chi}^2_{\rm min}+N^2$ for $N$ standard deviation ($\sigma$) uncertainties. 

The present experimental type B uncertainties have point-to-point correlations whose exact correlation matrix is 
difficult to evaluate precisely.
Thus, we consider two limiting correlation cases.
The first is where the type B uncertainties are 100\% 
correlated, i.e. all points move in the same direction by the same 
fraction of their respective type B uncertainty.  The second is where 
the type B uncertainties are correlated such that the low $p_{\rm T}$ and 
high $p_{\rm T}$ points may shift with opposite sign (and linearly scaled
in between), thus tilting the $R_{\rm AA}$ either upward or downward as a function of $p_{\rm T}$.
The minimum $\tilde{\chi}^2$ of the two cases is used for each constraint.

We take as a first example the resulting theory predictions from the PQM model.  For each calculation
characterized by the \qhatave, we calculate $\tilde{\chi}^2$.  We determine this
value by varying $\epsilon_b$ and $\epsilon_c$ (the systematic offsets) until we obtain the minimum  $\tilde{\chi}^2$.  
These values are shown in Fig.~\ref{fig_pqm_results} in the top panel.  
One can see 
that the overall lowest $\tilde{\chi}^2$ value corresponds to \qhatave~ $\approx 13$~GeV$^{2}$/fm.  

We then calculate the \pvalue~for the $\tilde{\chi}^2$ 
(the minimum of Eq.~\ref{eq:-2lnL-gaussian-sys}), where the
\pvalue~is defined as:

\begin{equation}
{\text {p-value}} = \int_{\tilde{\chi}^{2}}^{\infty} \chi^2_{(n_{d})}(z) dz
\end{equation}

\noindent
where $\chi^{2}_{(n_{d})}$ is the chi-square distribution with the 
appropriate number of degrees of freedom, $n_d$~\cite{PDG}.
This calculation is valid since the goodness-of-fit 
statistic $\tilde{\chi}^2$ follows a standard $\chi^{2}$ distribution.
Note that \pvalue~is the probability, under the assumption that the hypothesis is correct, 
of randomly obtaining data with a worse fit to the hypothesis than the experimental data under test~\cite{PDG}.
These \pvalue's are shown in the middle panel of Fig.~\ref{fig_pqm_results}.  


We find the overall minimum (or best) $\tilde{\chi}^2$, and 
then compute which \qhatave~scenarios are one and two standard deviations away from this minimum.
The PQM transport coefficient \qhatave~is constrained by the experimental data as 
13.2 $^{+2.1}_{-3.2}$ and  $^{+6.3}_{-5.2}$ ~GeV$^{2}$/fm at the one and two standard deviation levels, respectively.  
The two standard deviation constraints correspond to the 95\% confidence intervals.
We note that this range of large \qhatave~values is currently under intense theoretical debate (see
for example \cite{baier}).  Thus, the quoted \qhatave~constraint is for the model-dependent parameter of
the specific PQM implementation, and relating this parameter to the fundamental value of the mean transverse
momentum squared exchange per unit length traversed may substantially lower the value.

We apply the identical procedure to the GLV, WHDG, and ZOWW calculations and show 
those results in Figs.~\ref{fig_glv_results}, ~\ref{fig_whdg_results}, and ~\ref{fig_zoww_results}.  For the GLV calculations this results in a constraint of
$dN^{g}/dy = 1400^{+270}_{-150}$ and $^{+510}_{-290}$ at the one and two standard deviation levels, respectively.
Constraints for the WHDG model are $dN^{g}/dy = 1400^{+200}_{-375}$ and $^{+600}_{-540}$ at the
one and two standard deviation levels, respectively.
Constraints for the ZOWW model are $\epsilon_{0} = 1.9^{+0.2}_{-0.5}$ and $^{+0.7}_{-0.6}$ at the
one and two standard deviation levels, respectively.
All of these constraint results are summarized in Table~\ref{tab:result}.

For each of the above fits, there is a best fit value of $\epsilon_{b}$ and $\epsilon_{c}$ corresponding 
to the parameters in Eq.~\ref{eq:-2lnL-gaussian-sys}.  For completeness, these values are for PQM, GLV, WHDG, and ZOWW, 
$\epsilon_{b} = 0.6$ and $\epsilon_{c} = -0.3$,
$\epsilon_{b} = 0.7$ and $\epsilon_{c} = -0.0$,
$\epsilon_{b} = 2.1$ and $\epsilon_{c} = -1.5$, and
$\epsilon_{b} = 1.1$ and $\epsilon_{c} = -0.6$, respectively.  
All of the models considered here have a steeper \pt-dependence of $R_{AA}$ than the experimental data.
Thus, the best fit is obtained within the type B uncertainties by tilting the $R_{AA}$.

It is notable that although there is a well defined 
overall minimum in the modified $\tilde{\chi}^2$ for all
four models, the maximum \pvalue~in each case is different.  In the PQM, GLV, and ZOWW models the 
maximum \pvalue's are  approximately 9.0\%, 5.5\% and 7.8\%, respectively.  However, in the WHDG model the
maximum \pvalue~is substantially smaller at 1.3\%.  This is due to the fact that the WHDG model
has a steeper \pt-dependence of the nuclear modification factor $R_{\rm AA}$ (regardless of parameter
input) than the other models, and also steeper than the experimental data points.

The identical best value of $dN^{g}/dy$ for the GLV and WHDG calculations is interesting, since the inclusion
of important collisional energy-loss in WHDG leads to the naive expectation that a smaller color-charge
density would be needed for a similar suppression.   However, the different treatments of the distribution of paths 
through the medium may be compensating for this effect.  
Further theoretical studies are needed to disentangle the physics implications of the shape differences and the
similar best $dN^{g}/dy$ from GLV and WHDG.
For all the models considered, the relevant parameter constraint is approximately a $\pm 20-25$\% uncertainty at the one standard deviation level.

\begin{table*}[htbp]
 \caption{\label{tab:result}
Quantitative constraints on the model parameters from the PQM, GLV, 
WHDG, and ZOWW models and a linear functional form fit.}
\begin{ruledtabular} \begin{tabular}{llccc}
Model & Model & One Standard Deviation & Two Standard Deviation & Maximum \\
Name  & Parameter                  & Uncertainty            & Uncertainty & \pvalue \\
\hline
PQM   & \qhatave~= 13.2~GeV$^{2}$/fm & ${+2.1} \ \ {-3.2}$ & ${+6.3} \ \  {-5.2}$ & 9.0\% \\
GLV   & $dN^{g}/dy = 1400$ & ${+270} \ \  {-150}$ &  ${+510} \ \  {-290}$  & 5.5\% \\
WHDG  & $dN^{g}/dy = 1400$ & ${+200} \ \ {-375}$ & ${+600} \ \  {-540}$ & 1.3 \% \\
ZOWW  & $\epsilon_{0} = 1.9$~GeV/fm  & ${+0.2} \ \ {-0.5}$ & ${+0.7} \ \  {-0.6}$ & 7.8 \% \\
Linear & b (intercept) = $0.168$ & ${+0.033} \ \  {-0.032}$ & ${+0.065} \ \  {-0.066}$ & 11.6\% \\
       & m (slope) = $0.0017$ ($c$/GeV) & ${+0.0035 } \ \  {-0.0039}$  
       &  ${+0.0070} \ \ {-0.0076}$ & \\
\end{tabular} \end{ruledtabular}
\end{table*} 

\begin{figure}[tb]
\includegraphics[width=1.0\linewidth]{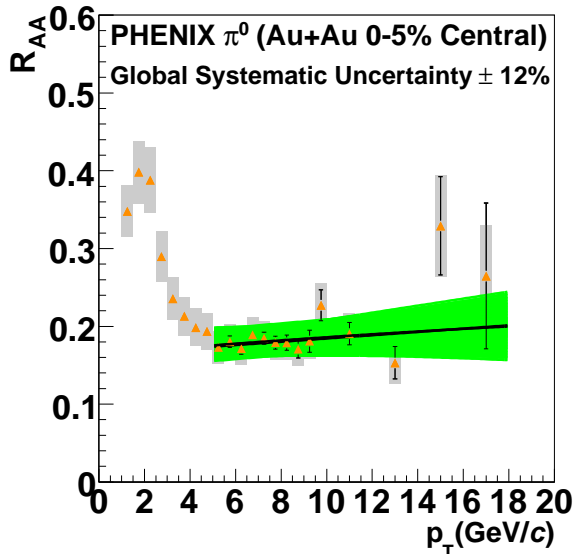}
\caption[]{
(Color online) The $\pi^{0}$ nuclear suppression factor $R_{\rm AA}$ as a function of 
transverse momentum for 0-5\% \auau collisions at $\sqrt{s_{NN}}$=200 GeV.  
Point-to-point uncorrelated statistical and systematic uncertainties are 
shown as uncertainty bars.  Correlated systematic uncertainties are shown as 
gray boxes around the data points.  The global scale factor systematic 
uncertainty is quoted as text. Also shown are the best fit and the 
envelope of lines with one standard deviation uncertainty for a simple 
linear fit function constrained by the statistical and systematic 
uncertainties.
}
\label{fig_simplefits_results}
\end{figure}

\begin{figure}[tb]
\includegraphics[width=1.0\linewidth]{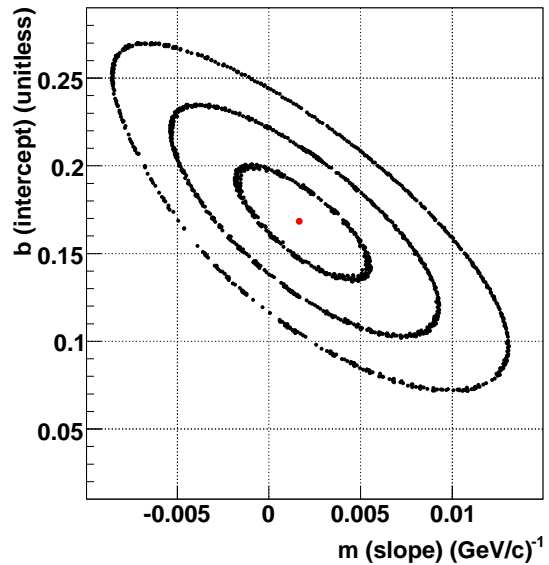}
\caption[]{
(Color online) Shown are the best fit values for m(slope) and b(intercept) as constrained by the experimental data.  
Also shown are the one, two and three standard deviation uncertainty contours.
}
\label{fig_simplefits_contour}
\end{figure}

It is also interesting to inquire what simple linear fit function best 
describes the experimental data for \pt $> 5$ GeV/$c$. The identical 
procedure to that described above is applied to the function 
$R_{\rm AA}(p_{\rm T}) = {\rm b} + {\rm m} \cdot $ \pt\
to determine the best values for the two 
parameters.  The best fit line and the envelope of lines with one standard 
deviation uncertainties are shown in Fig.~\ref{fig_simplefits_results}.  
The results including all types of uncertainties are ${\rm }b({\rm 
intercept}) = 0.168^{+0.033}_{-0.032}$ and ${\rm m}({\rm slope}) = 
0.0017^{+0.0035}_{-0.0039}$ ($c$/GeV).  The uncertainties on these 
parameters are correlated as shown by the one, two, and three standard 
deviation contours in Fig.~\ref{fig_simplefits_contour}.

Thus the data are consistent with a completely flat \pt-dependence of 
$R_{\rm AA}$ for \pt $> 5$ GeV/$c$ (i.e. $m = 0$) within one standard 
deviation uncertainties.  The maximum \pvalue~for this simple linear 
function fit is 11.6\%.

The \pvalue's for all models considered are less than 12\%.  It is notable 
that the five highest \pt ~points (\pt $> 9.5$ GeV/$c$), contribute over 
70\% to the total $\tilde{\chi}^{2}$.  As a check on the influence of 
these points on the extracted parameter values, we have repeated the above 
procedure to the restricted range 5 $<$ \pt $<$ 9.5~GeV/$c$.  We find the 
following new constraints: PQM model \qhatave~= 13.2 
$^{+1.8}_{-4.2}$~GeV$^{2}$/fm; GLV model $dN^{g}/dy = 1400^{+200}_{-210}$;  
WHDG model $dN^{g}/dy = 1000^{+300}_{-170}$; ZOWW model $\epsilon_{0} = 1.5 ^{+0.5}_{-0.2}$~GeV/fm;
simple linear fit ${\rm }b({\rm intercept}) = 0.170^{+0.034}_{-0.034}$ and ${\rm 
m}({\rm slope}) = 0.0013^{+0.0047}_{-0.0051}$ ($c$/GeV).  We find that 
the resulting new constraints are within the one standard deviation 
uncertainties of those quoted for the full \pt~range.  However, with 
the restricted range, the \pvalue's increase to 55\%, 36\%, 17\%, 62\%, and 75\% 
for the PQM model, GLV model, WHDG model, ZOWW model, and the simple linear fit, respectively.
Improvements in the data for $p_T > 9.5$ GeV/c expected from future
measurements will be crucial in determining whether any of the models discussed provide a
statistically valid description of the data over the full range $5\leq p_T\leq 20$ GeV/c.

\section{Summary and Conclusions}

In this paper, we have compared model predictions of parton energy-loss with 
experimental data of semi-inclusive single high transverse momentum 
$\pi^{0}$ suppression in central \auau reactions at $\sqrt{s_{NN}}$=200 
GeV.  In the comparison, statistical and systematic uncertainties were 
taken into account. We have obtained experimental constraints on model parameters 
of the color-charge density of the medium or its transport coefficient.  
These values indicate a large medium density.  
It is crucial to note that the quoted constraints on these parameters do 
not include any systematic uncertainties in the models, but rather give the limits 
assuming a ``perfect theory'' with one unknown parameter, for example the color-charge density, 
constrained by the measurements including the experimental 
statistical and systematic uncertainties. 
%
%
Additional theoretical systematic uncertainties from the time evolution, energy-loss approximations, and 
calculation details need further investigation.

\section{Acknowledgments}  

We thank the staff of the Collider-Accelerator and Physics Departments at 
Brookhaven National Laboratory and the staff of the other PHENIX 
participating institutions for their vital contributions.
We thank W. Horowitz, C. Loizides, I. Vitev, and X.-N. Wang for the 
theoretical calculation input and useful discussions.
We acknowledge support from the Office of Nuclear Physics in the Office of 
Science of the Department of Energy, the National Science Foundation, 
Abilene Christian University Research Council, Research Foundation of 
SUNY, and Dean of the College of Arts and Sciences, Vanderbilt University 
(U.S.A),
Ministry of Education, Culture, Sports, Science, and Technology
and the Japan Society for the Promotion of Science (Japan),
Conselho Nacional de Desenvolvimento Cient\'{\i}fico e
Tecnol{\'o}gico and Funda\c c{\~a}o de Amparo {\`a} Pesquisa do
Estado de S{\~a}o Paulo (Brazil),
Natural Science Foundation of China (People's Republic of China),
Ministry of Education, Youth and Sports (Czech Republic),
Centre National de la Recherche Scientifique, Commissariat
{\`a} l'{\'E}nergie Atomique, and Institut National de Physique
Nucl{\'e}aire et de Physique des Particules (France),
Ministry of Industry, Science and Tekhnologies,
Bundesministerium f\"ur Bildung und Forschung, Deutscher
Akademischer Austausch Dienst, and Alexander von Humboldt Stiftung (Germany),
Hungarian National Science Fund, OTKA (Hungary), 
Department of Atomic Energy (India), 
Israel Science Foundation (Israel), 
Korea Research Foundation and Korea Science and Engineering Foundation (Korea),
Ministry of Education and Science, Russian Academy of Sciences,
Federal Agency of Atomic Energy (Russia),
VR and the Wallenberg Foundation (Sweden), 
the U.S. Civilian Research and Development Foundation for the
Independent States of the Former Soviet Union, the US-Hungarian
NSF-OTKA-MTA, and the US-Israel Binational Science Foundation.

\appendix
\section{Constraint Formalism}


In the case of only point-to-point uncorrelated uncertainties (statistical 
and/or systematic), if one assumes they are Gaussian distributed and 
characterized by $\sigma$, the root-mean square (RMS), calculating the 
best-parameter fit is straightforward via a log-likelihood or least 
squares-$\chi^2$ method~\cite{PDG}.

The likelihood function $\cal L$ is defined as the \emph{a priori} 
probability of a given outcome.  Let $y_1, y_2, \ldots y_n$ be $n$ samples 
from a population with normalized probability density function 
$f(y,\vec{p})$ where $\vec{p}$ represents a vector of $k$ parameters. For 
instance $y_i$ could represent a measurement of a cross section at 
transverse momentum ($p_{\rm T})_i$, where the probability density of the 
measurement is Gaussian distributed about the expectation value 
$\mu=\mean{y}$:
 \begin{equation}
 f(y,\vec{p})={1\over \sqrt{2\pi\sigma^2}}\  \exp {- \left[ {(y-\mu)^2}\over {2\sigma^2} \right]}
 \qquad  
 \label{eq:gaussian}
 \end{equation}
If the samples are independent, then the likelihood function is:
\begin{widetext}
 \begin{equation}
 {\cal L}=\prod_i f(y_i,\vec{p})={1 \over {\sigma_1 \sigma_2 \ldots \sigma_n}}
 {1\over {\sqrt{2\pi}^n}}\  \exp {- \left[ \sum_{i=1}^{n} {(y_i-\mu_i(\vec{p}))^2 \over {2\sigma_i^2}}\right]}\qquad  
 \label{eq:L-gaussian}
 \end{equation}
However, if the samples are correlated, for example via correlated 
systematic uncertainties, then the full covariance matrix must be used 
  \begin{equation}
  V_{ij}=<(y_i -\mu_i(\vec{p}))(y_j-\mu_j(\vec{p}))> \qquad .
  \label{eq:Vij}
  \end{equation}
Then the likelihood function takes the more general form:
  \begin{equation}
   {\cal L}={1 \over {\sqrt{|V|}}} {1\over {\sqrt{2\pi}^n}}\ \exp {- \left[ \sum_{i=1}^{n} 
   \sum_{j=1}^{n} {{(y_i-\mu_i(\vec{p})) V^{-1}_{ij} (y_j-\mu_j(\vec{p}))}\over 2}\right] }   \qquad 
   \label{eq:Lcorr-gaussian}
  \end{equation}
where $|V|$ is the determinant of the covariance matrix $V$. Note that 
Eq.~\ref{eq:Lcorr-gaussian} reduces to Eq.~\ref{eq:L-gaussian} if the 
correlations vanish so that the covariances are zero and $V_{ij}$ is 
diagonal 
  \begin{equation}
  V_{ij}=<(y_i -\mu_i(\vec{p}))(y_j-\mu_j(\vec{p}))> =\delta_{ij} <(y_i -\mu_i(\vec{p}))^2>=\delta_{ij} \sigma_i^2 \qquad .
  \label{eq:V-diag}
 \end{equation}
Since Gaussian probability distributions are inevitable (as a consequence 
of the Central Limit Theorem) and since there is also an important theorem 
regarding likelihood ratios for composite hypotheses, it is convenient to 
use the logarithm of the likelihood
\begin{equation}
   -2 \ln {\cal L}={\ln{|V|}}+ {n\log{2\pi}} + {\sum_{i=1}^{n} 
   \sum_{j=1}^{n} {{(y_i-\mu_i(\vec{p})) V^{-1}_{ij} (y_j-\mu_j(\vec{p}))}} }   \qquad .
   \label{eq:-2lnLcorr-gaussian}
\end{equation}
\end{widetext}

We separate the uncertainties into four classes: type A) point-to-point 
uncorrelated systematic uncertainties; type B) correlated systematic 
uncertainties, for which the point-to-point correlation is 100\% by 
construction, since the uncorrelated part has been separated out and 
included in uncertainty A); type C) overall systematic uncertainties by 
which all the points move by the same fraction (i.e. normalization 
uncertainties); and type D) statistical.  Categories A and D are simply added in 
quadrature and represent the total point-to-point uncorrelated 
uncertainties, denoted $\sigma_{i}$ below.

We model a correlated systematic uncertainty as if there were an 
underlying uncertainty, e.g. absolute momentum scale, which may cause 
correlated systematic variations $\Delta y_{i}$ of the set of 
measurements, $y_{i}$, around their nominal value, that can be represented 
as a random variable, $z_b$. The correlated type B variation of the 
measurements is represented by the displacement of all points from their 
nominal values by the correlated amounts
\begin{eqnarray}
     \Delta y^{b}(sys)_i&\equiv&b_i \Delta z_b\label{eq:eb}\\
  \mean{\Delta y^{b} (sys)_i\, \Delta y^{b} (sys)_j}&=&b_i b_j \sigma^2_b\equiv \sigma_{b_i}\sigma_{b_j}\label{eq:veb},
\end{eqnarray}
\noindent
where $\Delta z_b\equiv z_b-\langle z_b\rangle=z_b$.

Since $\mean{z_b}\equiv 0$, $\langle (\Delta z_b)^2\rangle=\langle 
z_b^2\rangle-\langle z_b\rangle^2=\sigma_b^2$ and the random variable 
$z_b$ is the same for all $i$ measurements while $b_i$ is a constant of 
proportionality which may be different for each $i$.
We define $\sigma_{b_i}\equiv b_i \sigma_b$, where $\pm|\sigma_{b_i}|$ is 
the systematic uncertainty bar shown on each point (gray box on each data point 
in Fig.~\ref{fig_data_only} ) and where $b_i$ may be of either sign, as 
it is possible that one point could move up while its neighbor moves down.  
The random variable $z_b$ is assumed to have a Gaussian probability 
distribution $f({z_b})$, with r.m.s. $\sigma_b$
\begin{equation} 
   f(z_b)=\frac{1}{\sigma_b \sqrt{2\pi}} \exp {- \left[ {(\Delta z_b)^2}\over {2\sigma_b^2} \right]} \qquad . 
   \label{eq:gausszb}
\end{equation}
  
The type C variation is independent of the type B variation. It is 
similarly assumed to be caused by an underlying random variable $z_c$ that 
results in a systematic displacement of the measurement by an amount 
$\Delta y_i^c$ with
\begin{eqnarray}
\Delta y^{c}(sys)_i/y_i&\equiv&\Delta z_c \label{eq:ec}\\
\mean {(\Delta y^{c}(sys)_i/y_i)\, (\Delta y^{c}(sys)_j/y_j)}&=&\sigma_c^2 \label{eq:vec}
\end {eqnarray}
\noindent
where by definition $\sigma_c$ is the same for all points. 
 
We then assume that the Likelihood function factorizes as the product of 
independent Gaussian probabilities as in Eq.~\ref{eq:L-gaussian}, but that 
the distributions are correlated through their dependence on the random 
variables $z_b$ and $z_c$:
\begin{widetext}
\begin{eqnarray}
{\cal L}&=&\prod_i f(y_i;z_b,z_c,\vec{p})f(z_b)f(z_c)  \\
&=&   {1 \over {\sigma_1 \sigma_2 \ldots \sigma_n \sigma_b \sigma_c}}
  {1\over {\sqrt{2\pi}^{(n+2)}}}\  \exp \left\{ - \left[ \left( \sum_{i=1}^{n} 
      {{(y_i+b_i \Delta z_b + y_i \Delta z_c -\mu_i(\vec{p}))^2} \over {2\sigma_i^2}} \right) + {(\Delta z_b)^2 \over {2\sigma_b^{2}}}+{(\Delta z_c)^2 \over {2\sigma_c^{2}}}\right] \right\}  \qquad .
\label{eq:L-witherrors}
\end {eqnarray}
\noindent

To account for the type B systematic uncertainty, we allow any given 
sample of measurements, $y_i$, corresponding to theoretical predictions 
$\mu_i(\vec{p})$ to have a correlated variation from their nominal values 
by an amount corresponding to a certain fraction $\epsilon_b$ of the 
underlying root-mean-square variation of $z_b$, i.e. $\Delta 
{z_b}=\epsilon_b \sigma_b$, such that each point moves by an amount 
$\Delta y_i^b=b_i \epsilon_b \sigma_b \equiv \epsilon_b \sigma_{b_i}$, the 
same fraction $\epsilon_b$ of its systematic uncertainty bar; and similarly for 
the type C uncertainty. Then the likelihood function for any outcome, 
including the variation of $\epsilon_b$ and $\epsilon_c$ would be:

\begin{equation}  
     {\cal L}={1 \over {\sigma_1 \sigma_2 \ldots \sigma_n \sigma_b \sigma_c}}
     {1\over {\sqrt{2\pi}^{(n+2)}}}\  \exp \left\{ - \left[ \left( \sum_{i=1}^{n} 
     {{(y_i+\epsilon_b \sigma_{b_i} +\epsilon_c y_i \sigma_c -\mu_i(\vec{p}))^2} \over {2\sigma_i^2}} \right) + {\epsilon_b^2 \over {2}}+{\epsilon_c^2 \over {2}}\right] \right\}
     \label{eq:L-gaussian-sys}
\end{equation}
\noindent
where the last two terms represent $(\Delta z_b)^2 / (2 \sigma^2_b)=\epsilon^2_b\sigma_b^2 / (2 \sigma^2_b)$ and
$(\Delta z_c)^2 / (2 \sigma^2_c)=\epsilon^2_c\sigma_c^2 / (2 \sigma^2_c)$ 
since we assumed the probability of the systematic displacements $f(z_{b,c})$ to be Gaussian. Other probability distributions for the correlated systematic uncertainty could be used. For instance if $\pm|\sigma_{b_i}|$ had represented full extent systematic uncertainties, with equal probability for any $\Delta {z_b}$, then the $\epsilon_b^2/2$ term and associated normalization constant $1/\sigma_b\sqrt{2\pi}$ would be absent from Eq.~\ref{eq:L-gaussian-sys}.  

Then we use the likelihood ratio test to establish the validity or the 
confidence interval of the theoretical predictions $\mu_i(\vec{p})$. One 
can use the modified log likelihood

\begin{equation}
-2\ln {\cal L}={\left[ \left( \sum_{i=1}^{n}
{{(y_i+\epsilon_b \sigma_{b_i} +\epsilon_c y_i \sigma_c -\mu_i(\vec{p}))^2}  \over {\sigma_i^2}} \right) + {\epsilon_b^2 }+{\epsilon_c^2 }\right]}\equiv \chi^2(\epsilon_{b}, \epsilon_{c}, \vec{p})
\label{eq:-2lnL-gaussian-sys2}
\end{equation}
\end{widetext}
\noindent
because we will eventually take the ratio of the likelihood of a given set 
of parameters $\vec{p}$ to the maximum likelihood when all the parameters 
$\epsilon_b$, $\epsilon_c$ and $\vec{p}$ are varied (the minimum value of 
Eq.~\ref{eq:-2lnL-gaussian-sys2}) so that the terms preceding the 
exponential in Eq.~\ref{eq:L-gaussian-sys} cancel because they are not 
varied. 
Eq.~\ref{eq:-2lnL-gaussian-sys2} follows the $\chi^2$-distribution with 
$n+2$ degrees of freedom because it is the sum of $n+2$ independent 
Gaussian distributed random variables (i.e. in statistical terminology 
$\chi^2(\epsilon_{b}, \epsilon_{c}, \vec{p})$ is $\chi^2_{(n+2)}$ ). This 
establishes Eq.~\ref{eq:-2lnL-gaussian-sys2} as the $\chi^2$-distributed 
quantity that we use for least squares fit to the theoretical predictions 
including the systematic uncertainties. Note that 
Eq.~\ref{eq:-2lnL-gaussian-sys2} agrees with Eq.~8 in~\cite{heinrich} 
in the discussion of fits with correlated systematics. The specific 
procedure is described in the next paragraph.

First `fit the theory' to the data by minimizing 
Eq.~\ref{eq:-2lnL-gaussian-sys2} by varying all the parameters to find 
$\hat{\epsilon}_b$, $\hat{\epsilon}_c$, $\vec{\hat{p}}$, the values of the 
parameters which give the overall minimum $\chi^2_{\rm min}$. If the 
$\chi^2_{\rm min}$ for this fit is acceptable for the $n+2-(m+2)=n-m$ 
degrees of freedom, where $m$ are the number of parameters in $\vec{p}$, 
then the theory is not rejected at this level. A confidence interval is 
then found for testing any other set of $k$ parameters constrained to 
specific values, $\vec{p}_0$, by again finding the minimum of 
Eq.~\ref{eq:-2lnL-gaussian-sys2} for the $k$ fixed values of $\vec{p}_0$, 
by letting all the other parameters including $\epsilon_b$ and 
$\epsilon_c$ vary.  For constant values of $\sigma_i$, and large values of 
$n$, the ``likelihood ratio'' $-2\ln [{\cal L}(\vec{p}_0)/{\cal 
L}(\vec{\hat{p}})]=-2[\ln{\cal L}(\vec{p}_0)-\ln{\cal L}(\vec{\hat{p}})]$, 
i.e. $\chi^2(\vec{p}_0)-\chi^2_{\rm min}$ is $\chi^2$-distributed with $k$ 
degrees of freedom, from which the confidence interval on the parameters 
can be evaluated. However, in general, the uncertainty on the parameters 
is estimated in the Gaussian approximation by 
$\chi^2(\vec{p}_0)=\chi^2_{\rm min}+N^2$ for $N$ standard deviation 
uncertainties (for example using a MINUIT~\cite{MINUIT} type fitting 
algorithm).

For the present data, the statistical and random systematic uncertainties 
are such that the shift in the measurement $y_i$ due to the correlated 
systematic uncertainties preserves the fractional uncertainty. In this 
case the maximum likelihood and least squares methods no longer coincide 
and we use a least squares fit of Eq.~\ref{eq:lstsq-appendix} instead of 
Eq.~\ref{eq:-2lnL-gaussian-sys2} to estimate the best fit parameters:  
\begin{equation}
\tilde{\chi}^2={\left[ \left( \sum_{i=1}^{n}
{{(y_i+\epsilon_b \sigma_{b_i} +\epsilon_c y_i \sigma_c -\mu_i(\vec{p}))^2}  \over {{\tilde{\sigma}}_i^2}} \right) + {\epsilon_b^2 }+{\epsilon_c^2 }\right]} \qquad , 
\label{eq:lstsq-appendix}
\end{equation}
where ${\tilde{\sigma}}_{i}$ is the uncertainty scaled by the multiplicative shift in $y_i$ such that the
fractional uncertainty is unchanged under shifts
\begin{equation}
\tilde{\sigma}_i=\sigma_i \left( \frac{y_i+\epsilon_b \sigma_{b_i} +\epsilon_c y_i \sigma_c}{y_i}\right) \qquad . 
\label{eq:tildesigma}
\end{equation}


\end{document}